\documentclass[12pt]{article}
\usepackage{graphicx,amsfonts,cite}
\usepackage{geometry}
\begin{document}

\title{Confluent Second-Order Supersymmetric Quantum Mechanics and Spectral Design}
\author{{\sl David J. Fern\'andez C.}\\Departamento de F\'isica, Cinvestav \\ A.P. 14-740, 07000 Ciudad de M\'exico, Mexico\\david@fis.cinvestav.mx\\[12pt]{\sl Barnana Roy}\\Physics \& Applied Mathematics Unit\\ Indian Statistical Institute\\ Kolkata - 700 108, India\\taturoy@gmail.com}
\date{}	
\maketitle
\begin{abstract}
The confluent second-order supersymmetric quantum mechanics, in which the factorization energies tend to a common value, is used to generate Hamiltonians with known spectra departing from the hyperbolic Rosen-Morse and Eckart potentials. The possible spectral modifications, as to create a new level or to delete a given one, as well as the isospectral transformations, are discussed.
\end{abstract}

\section{Introduction}
Supersymmetric quantum mechanics (SUSY QM) \cite{cooper,junker,bagchi} is a powerful tool for building Hamiltonians with a prescribed spectrum in quantum physics \cite{andri1,andri2,fern20,fern10,fern19}. This technique, together with the factorization method \cite{infeld} and Darboux transformation \cite{darboux}, are equivalent procedures for generating solvable Hamiltonians with a modified spectrum departing from an initial solvable one.
The underlying idea of these procedures is an algebraic scheme known as intertwining approach. Its simplest version, the so-called first-order SUSY QM, involves the intertwining operators which are first order differential ones, but with  
the restriction that spectral modification can be done only below the initial ground state energy level, in order to avoid singularities in the constructed SUSY partner potential. On the other hand, the higher-order SUSY, in particular the second order one, offers several interesting possibilities of spectral modification \cite{fern20}. The ingredients to implement this transformation are solutions, termed as seed solutions, to the Schr\"odinger equation of the initial system for different factorization energies, which do not coincide in general with the eigenvalues of the initial Hamiltonian. The transformed system is then characterized by the Wronskian of the chosen seed solutions. This standard algorithm has got its confluent counterpart, known as confluent SUSY algorithm \cite{Mielnik00}. In the confluent SUSY QM of higher order the seed solutions satisfy a coupled system of differential equations, often referred to as Jordan chain, and several factorization energies converge to the same value. The mathematical properties of the Wronskian in confluent second and higher-order SUSY can be found in \cite{fern1,fern2,fern3,berm1,berm2,halberg2,conte4}, while the Jordan chain analysis is given in \cite{halberg3,conte2,conte1,halberg5}.\\
The confluent SUSY techniques have been applied to several examples, for which the $x$-domain is either the full real line (e.g., the harmonic oscillator \cite{fern1}), a finite interval (the trigonometric P\"oschl Teller potental \cite{conte10}) or the positive semi-axis (e.g. the radial oscillator or the Coulomb problem \cite{fern2}). They have been implemented also for the one-gap Lame potential in \cite{fern1,fern2,fern3,berm1,berm2,conte2}. The application of the former to Dirac equation and  $\cal{PT}$-symmetric systems can be found in \cite{halberg10,correa1} and \cite{correa2} respectively. In \cite{gran} the confluent chains of Darboux-Backlund transformations have been applied for generating rational extensions of P\"oschl-Teller and isotonic oscillator potentials. In addition, the extended potentials possess an enlarged shape invariance \cite{genden1,genden2} and are associated to new families of exceptional orthogonal polynomials \cite{ullate1,ullate2}. As far as we know, however, the confluent second-order algorithm has not been used to implement spectral modification for shape invariant potentials involving hyperbolic functions, whose bound state solutions are associated to orthogonal polynomials. This motivated us to look into the spectral modification of the following two potentials through confluent second order SUSY QM: the hyperbolic Rosen-Morse potentials
$$ V(x) = -s(s+1){\rm sech}^2 (x) + 2\Lambda \tanh (x), \quad -\infty<x<\infty,$$
and the Eckart potential
$$ V(x) = A(A-1){\rm cosech}^2(x) - 2B \coth(x),\quad 0<x<\infty,$$ where $s$, $\Lambda$, $A$, $B$ are real parameters such that $s>0, \ 0<\Lambda<s^2$ 
and $A>1, \ B>A^2$.\\
The paper is organized as follows. In Section 2, we shall introduce and discuss the second-order confluent SUSY QM. In Section 3 we shall present two applications of the results obtained in Section 2. Finally, our conclusions are presented in the last section. 

\section{Second-order SUSY QM}
In the second order SUSY QM, the starting point is the following intertwining relationship between two Schr\"odinger Hamiltonians $H$ and $\tilde{H}$
\begin{equation}\label{eq1}
\tilde{H} B^+ = B^+ H,
\end{equation}
\begin{equation}\label{eq2}
H = - \frac{d^2}{dx^2} + V(x),
\end{equation}
\begin{equation}\label{eq3}
\tilde{H} = - \frac{d^2}{dx^2} + \tilde{V}(x),
\end{equation}
\begin{equation}\label{eq4}
B^+ = \frac{d^2}{dx^2} + \eta(x) \frac{d}{dx} + \gamma(x),
\end{equation}
where $\tilde{V}(x), \ \eta(x)$ and $\gamma(x)$ can be determined if the initial potential $V(x)$ is known.
The substitution of Eqs.~(\ref{eq2}-\ref{eq4}) in (\ref{eq1}) yields 
\begin{equation}\label{eq5}
\tilde{V} = V + 2 \eta',
\end{equation}
\begin{equation}\label{eq6}
\gamma = d - V + \frac{\eta^2}{2} - \frac{\eta'}{2},
\end{equation}
\begin{equation}\label{eq7}
\frac{\eta \eta''}{2} - \frac{(\eta')^2}{4} + \eta^2 \left(\frac{\eta^2}{4} -  \eta' - V +d\right) + c = 0,
\end{equation}
with $ c,d \in \mathbb{R}$ being two integration constants. Thus the problem defined by Eqs.(\ref{eq1}-\ref{eq4}) reduces to find the function $\eta(x)$. In order to solve (\ref{eq7}), the following ansatz is used
\begin{equation}\label{eq8}
\eta'(x) = \eta^2(x) + 2\eta(x) g(x) - 2\xi(x),
\end{equation}
where $g(x)$ and $\xi(x)$ are functions to be determined. Substitution of Eq.~(\ref{eq8}) in (\ref{eq7}) gives $ \xi^2 = c$ and the following Ricatti equation 
\begin{equation}\label{eq9}
g' + g^2 = V - \epsilon,
\end{equation}
with $\epsilon = d + \xi$. This equation can be linearized by using $g = u'/u$, which leads to the following stationary Schr\"odinger equation for $H$
\begin{equation}\label{eq10}
H u = -u'' + V u = \epsilon u.
\end{equation}
The kind of seed solution $u$ employed for constructing the SUSY transformation depends on the factorization energy $\epsilon$ and, consequently, on the sign of $c$. For $c\neq 0$ one gets the real and complex cases, while for $c = 0$ the confluent case is obtained.\\

\subsection{Non-confluent case ($c\neq0$)} 
In this case there are two different factorization energies given by $\epsilon_1 = d + \sqrt{c}$ and $\epsilon_2 = d - \sqrt{c}$. The real case corresponds to $c>0$, $\epsilon_1,\epsilon_2 \in \mathbb{R}$ and the complex one to $c<0$, $\epsilon_1 \in \mathbb{C}, \epsilon_2 = \epsilon_1^*$. The ansatz (\ref{eq8}) gives rise to the following two equations
\begin{equation}\label{eq11}
\eta' = \eta^2 + 2 g_1 \eta - (\epsilon_1 - \epsilon_2),
\end{equation}
\begin{equation}\label{eq12}
\eta' = \eta^2 + 2 g_2 \eta + (\epsilon_1 - \epsilon_2).
\end{equation}
Subtraction of the above two equations gives
\begin{equation}\label{eq13}
\eta = \frac{\epsilon_1 - \epsilon_2}{g_1 - g_2} = -\{\ln[W(u_1,u_2)]\}' ,
\end{equation}
where $W(f,h) = fh'-f'h$ is the Wronskian of $f$ and $h$. In order to avoid singularities in $\eta$ and, consequently in $\tilde{V}$, the Wronskian in (\ref{eq13}) should not have zeros. Substituting (\ref{eq13}) in (\ref{eq5}) leads to
\begin{equation}\label{eq14}
\tilde{V} = V - 2\{\ln[W(u_1,u_2)]\}'' .
\end{equation}
This expression has been used to construct new potentials $\tilde{V}$ departing from the initial one $V$ by choosing two appropriate solutions of (\ref{eq10}). Several possible spectral modification include: (i) creation of two levels below the ground state energy of $H$; (ii) insertion of a pair of levels between two neighbouring energies of $H$; (iii) Deletion of two neighbouring energies of $H$; (iv) isospectral transformations; (v) generation of complex potentials with real spectra \cite{fern20,fern1,fern2,fern3,conte10}.

\subsection{Confluent case ($c=0$)}
In this case $\epsilon_1 = \epsilon_2 \equiv \epsilon$. Consequently the ansatz (\ref{eq8}) becomes
\begin{equation}\label{eq15}
\eta' = \eta^2 + 2 g \eta .
\end{equation}
The general solution of this Bernoulli equation is given by
\begin{equation}\label{eq16}
\eta(x) = \frac{e^{2\int g(x)dx}}{w_0 - \int e^{2\int g(x)dx }dx} = -\{\ln[w(x)]\}' ,
\end{equation}
where 
\begin{equation}\label{wofx}
w(x) = w_0 - \int_{x_0}^x u^2(y) dy ,
\end{equation}
with $w_0\equiv w(x_0)$ being a real constant which can be chosen appropriately to avoid singularities in $\eta(x)$. 
The confluent second order SUSY partner potential $\tilde{V}(x)$ is now given by:
\begin{equation}\label{tildeV}
\tilde{V}(x) = V(x) - 2\{\ln[w(x)]\}'' .
\end{equation}
Note that $w(x)$ must be nodeless in order that $\tilde{V}(x)$ has no singularities in $(-\infty, \infty)$. As
\begin{equation}
w'(x) = -[u(x)]^2 ,
\end{equation}
i.e., $w(x)$ is a monotonic decreasing function, thus one has to look for the appropriate asymptotic behaviour for $u(x)$ in order to avoid one possible zero in $w(x)$. The following two different cases arise here \cite{fern1}:\\
(i) Let us suppose that $\epsilon = E_n$ is one of the discrete eigenvalues of $H$ and the seed solution is the corresponding normalized physical eigenfunction $u(x) = \psi_n(x)$.
Let us denote by $\mu_+$ the following integral:
\begin{equation}
\mu_+ \equiv \int_{x_0}^{\infty} u^2(y)dy .
\end{equation}
Then, it is straightforward to show that
\begin{equation}
\lim_{x\rightarrow -\infty} w(x) = w_0 - \mu_+ + 1 \equiv \mu + 1 ,
\end{equation}
where $w_0 - \mu_+ = \mu$, while
\begin{equation}
\lim_{x\rightarrow \infty} w(x) = \mu .
\end{equation}
It  turns out that $w(x)$ is nodeless if either both limits are positive or both negative, thus the $\mu$-domain where the confluent second-order SUSY transformation is non-singular becomes:
\begin{equation}
\mu \in \mathbb{R} \backslash (-1,0) = (-\infty, -1] \cup [0,\infty) .
\end{equation}
(ii) If the transformation function $u(x)$ is a non-normalizable solution of (\ref{eq10}) for a real factorization energy $\epsilon$ which does not belong to the spectrum of $H$ ($\epsilon \notin {\rm Sp}(H)$) such that 
\begin{equation}
\lim_{x\rightarrow \infty} u(x) = 0~~ {\rm and} ~~\mu_+ = \int_{x_0}^{\infty} u^2(y) dy < \infty ,
\end{equation}
then it can be shown that 
\begin{equation}
\lim_{x \rightarrow -\infty} w(x) = w_0 + \int_{-\infty}^{x_0} u^2(y) dy = \infty ,
\end{equation}
while
\begin{equation}
\lim_{x \rightarrow +\infty} w(x) = w_0 - \mu_+ \equiv \mu .
\end{equation}
By comparing both limits, and keeping in mind that $w(x)$ is monotonic decreasing, it turns out that $w(x)$ is nodeless if
\begin{equation}
\mu \geq 0 .
\end{equation}
It is to be noted that the same $\mu$-restriction holds in the case when
\begin{equation}
\lim_{x\rightarrow -\infty} u(x) = 0~~ {\rm and}~~ \mu_- \equiv \int_{-\infty}^{x_0} u^2(y)dy < \infty ,
\end{equation}
though now $\mu \equiv -(w_0 + \mu_-)$.

\section{Applications}
Let us use now Eq.~(\ref{tildeV}) to implement a confluent second order SUSY transformation for two interesting systems. \\
1. {\bf Rosen Morse II Potential.}
Two linearly independent solutions $\psi^{\pm}$ of the Schr\"odinger equation 
\begin{equation}
\frac{d^2 \psi}{dx^2} + [\epsilon - V(x)]\psi = 0 ,
\end{equation}
with the Rosen-Morse II potential
\begin{equation}
V(x) = -\frac{s(s+1)}{\cosh^2(x)} + 2 \Lambda \tanh(x) ,
\end{equation}
can be written in terms of the hypergeometric function ${}_2F_{1}(a,b,c,z)$ \cite{abra} as follows \cite{levai}:
\begin{eqnarray}
&& \hskip-1.0cm {\psi}^{+} \! \approx \! (1\!-\!\tanh x)^{\frac{\alpha}{2}}~ (1\!+\!\tanh x)^{\frac{\beta}{2}} \, {}_2F_{1}\left(\!-s \!+\! \frac{\alpha \!+\! \beta}{2}, s\!+\!1\!+\!\frac{\alpha\!+\!\beta}{2}, 1\!+\!\alpha, \frac{1\!-\!\tanh x}{2}\right), \\
&& \hskip-1.0cm {\psi}^{-} \! \approx \! (1\!-\!\tanh x)^{\frac{-\alpha}{2}}~ (1+\tanh x)^{\frac{\beta}{2}} \, {}_2F_{1}\left(\!-s \!+\! \frac{\beta \!-\!\alpha}{2}, s\!+\!1\!+\!\frac{\beta\!-\!\alpha}{2}, 1\!-\!\alpha, \frac{1\!-\!\tanh x}{2}\right),
\end{eqnarray}
where $\alpha$ and $\beta$ are two new parameters, specified in terms of $\epsilon$ and $\Lambda$ as
\begin{equation}
\Lambda + \left(\frac{\beta - \alpha}{2}\right)\left(\frac{\alpha + \beta}{2}\right) = 0, ~~~~\epsilon + \left(\frac{\alpha + \beta}{2}\right)^2 + \left(\frac{\beta - \alpha}{2}\right)^2 = 0 .
\end{equation}
The Schr\"odinger seed solutions $u(x)$ with the right asymptotic behavior are constructed departing from the general solution
\begin{equation}
u(x) = D^+ \psi^+(x,\epsilon) + D^- \psi^-(x,\epsilon).
\end{equation}
By using the asymptotic properties of the hypergeometric function it turns out that:
\begin{equation}
u(x)(x\rightarrow\infty) \approx D^+ \, 2^{\frac{\alpha+\beta}{2}} e^{-\alpha x} + D^- \, 2^{\frac{\beta-\alpha}{2}} e^{\alpha x} . 
\end{equation}
Therefore, the condition $\lim\limits_{x\rightarrow\infty} u(x) = 0$ will be satisfied if $D^-$ is set equal to zero. In such a case the seed solution with the proper behavior for $x\rightarrow\infty$ is given by (taking $D^+=1$):
\begin{equation}\label{urightnullrm}
\hskip-0.2cm u_{1} \!=\! (1 - \tanh x)^{\frac{\alpha}{2}} (1 + \tanh x)^{\frac{\beta}{2}} \, {}_2F_{1}\left(-s \!+\! \frac{\alpha \!+\! \beta}{2}, s\!+\!1\!+\!\frac{\alpha\!+\!\beta}{2}, 1\!+\!\alpha, \frac{1\!-\!\tanh x}{2}\right) .
\end{equation}
Likewise, the seed solution with proper behavior for $x\rightarrow-\infty$ is given by (taking $D^+=1$):
\begin{eqnarray}
&& \hskip-0.4cm u_{2} = (1+\tanh x)^{\frac{\beta}{2}} \bigg[(1-\tanh x)^{\frac{\alpha}{2}} \, {}_2F_1\left(-s + \frac{\alpha + \beta}{2}, s+1+\frac{\alpha+\beta}{2}, 1+\alpha, \frac{1-\tanh x}{2}\right) \nonumber \\
&& \hskip0.6cm -  D^-
(1-\tanh x)^{-\frac{\alpha}{2}} \, {}_2F_1\left(-s + \frac{\beta -\alpha}{2}, s+1+\frac{\beta-\alpha}{2}, 1-\alpha, \frac{1-\tanh x}{2}\right)\bigg],
\label{uleftnullrm}
\end{eqnarray}
where 
$$ 
D^- = 2^{\alpha}\frac{\Gamma (1+\alpha) \Gamma (-s + \frac{\beta - \alpha}{2}) \Gamma (s + 1 + \frac{\beta - \alpha}{2})}{\Gamma (1-\alpha) \Gamma (-s + \frac{\beta + \alpha}{2}) \Gamma (s + 1 + \frac{\beta + \alpha}{2})}, \quad \alpha = \sqrt{-\epsilon + 2\Lambda}, \quad \beta^2 = -(\epsilon + 2\Lambda) ,
$$
and $\Gamma(x)$ is the Gamma function \cite{abra}. Thus, $\alpha$ and $\beta$ will be real for $2\Lambda > \epsilon$ and $\epsilon + 2\Lambda <0$, respectively.\\
The eigenfunctions fulfilling the boundary conditions 
\begin{equation}
\lim_{x\rightarrow\pm\infty} \psi_n(x) = 0 ,
\end{equation}
can be expressed in terms of the Jacobi polynomials as 
\begin{equation}\label{boundstatesrm}
\psi_n(x) = c_n (1-\tanh x)^{\frac{\alpha_n}{2}} (1+\tanh x)^{\frac{\beta_n}{2}} P_n^{(\alpha_n,\beta_n)}(\tanh x),
\end{equation}
where $c_n$ is the normalization constant and
\begin{eqnarray}
&& \alpha_n = s - n + \frac{\Lambda}{s-n}, \qquad \beta_n = s - n - \frac{\Lambda}{s-n}, \\
&& \hskip1.5cm E_n = -(s-n)^2 - \frac{\Lambda^2}{(s-n)^2}.
\end{eqnarray}
{\bf Spectral design}\\
(a) Isospectral transformation\\
The confluent isospectral transformation can be generated by using as seed solution a normalized physical eigenfunction of
$H$, i.e., by taking $\epsilon = E_n$ and $u(x) = \psi_n(x)$ (see Eq.~(\ref{boundstatesrm})). 
The integral in Eq.~(\ref{wofx}) with $x_0=-\infty$ can be obtained analytically, leading to
\begin{equation}\label{donewofx}
\int_{-\infty}^{x} \psi_n^2(y)dy = \frac{S_n(x)}{S_n(\infty)} ,
\end{equation}
where
\begin{equation}
\begin{array}{l}
\hskip-0.0cm S_n(x) \!=\!\! \sum\limits_{m'=0}^n \left(\!\!\!\begin{array}{c} n+\alpha_n\\m'\end{array}\!\!\!\right)\!\left(\!\!\!\begin{array}{c} n+\beta_n\\n-m'\end{array}\!\!\!\right) \frac{1}{(-1)^{m'} 2^{m'}} \sum\limits_{m=0}^n \left(\!\!\!\begin{array}{c} n+\alpha_n\\m\end{array}\!\!\!\right)\!\left(\!\!\!\begin{array}{c} n+\beta_n\\n-m\end{array}\!\!\!\right) \frac{1}{(-1)^{m} 2^{m}} \frac{1}{\beta_n+m+m'}\\
\hskip1.5cm (1\!+\!\tanh x)^{\beta_n+m+m'} \, {}_2F_1(\beta_n\!+\!m\!+\!m',-\alpha_n\!-\!2n\!+\!m\!+\!m'\!+\!1,\beta_n\!+\!m\!+\!m'\!+\!1,\frac{1+\tanh x}{2}),
\end{array} 
\end{equation}
\begin{equation}
\begin{array}{l}
\hskip-0.0cm S_n(\infty) \!= 2^{\beta_n} \sum\limits_{m'=0}^n \left(\!\!\!\begin{array}{c} n+\alpha_n\\m'\end{array}\!\!\!\right)\!\left(\!\!\!\begin{array}{c} n+\beta_n\\n-m'\end{array}\!\!\!\right) \frac{1}{(-1)^{m'}} \sum\limits_{m=0}^n \left(\!\!\!\begin{array}{c} n+\alpha_n\\m\end{array}\!\!\!\right)\left(\!\!\!\begin{array}{c} n+\beta_n\\n-m\end{array}\!\!\!\right) \frac{1}{(-1)^{m}} \\
\hskip3.4cm \frac{1}{\beta_n+m+m'} \frac{\Gamma (\beta_n + m + m'+1) \Gamma(\alpha_n+2n-m-m')}{\Gamma(\alpha_n+\beta_n+2n)},
\end{array}
\end{equation}
with $\Gamma(x)$ and $\left(\begin{array}{c} p\\q\end{array}\right)$ denoting the Gamma function and Binomial coefficient respectively \cite{abra}.\\
The confluent 2-SUSY partner potential can be obtained by plugging Eq.~(\ref{donewofx}) into Eqs.~(\ref{wofx}) and (\ref{tildeV}). Since the analytical expression is too cumbersome, we refrain here from giving the same. Instead, from such expression we have plotted the final potential (black continuous line) and compare it with the initial potential (blue dashed curve) in Figure 1. We have shown as well the first few energy levels of the common spectrum (red dotted horizontal lines). Let us note that the numerical calculation of $w(x)$ produces the same plot for the final potential as our analytic formulas.
\begin{figure}
	\centering
	\includegraphics[width=0.8\textwidth]{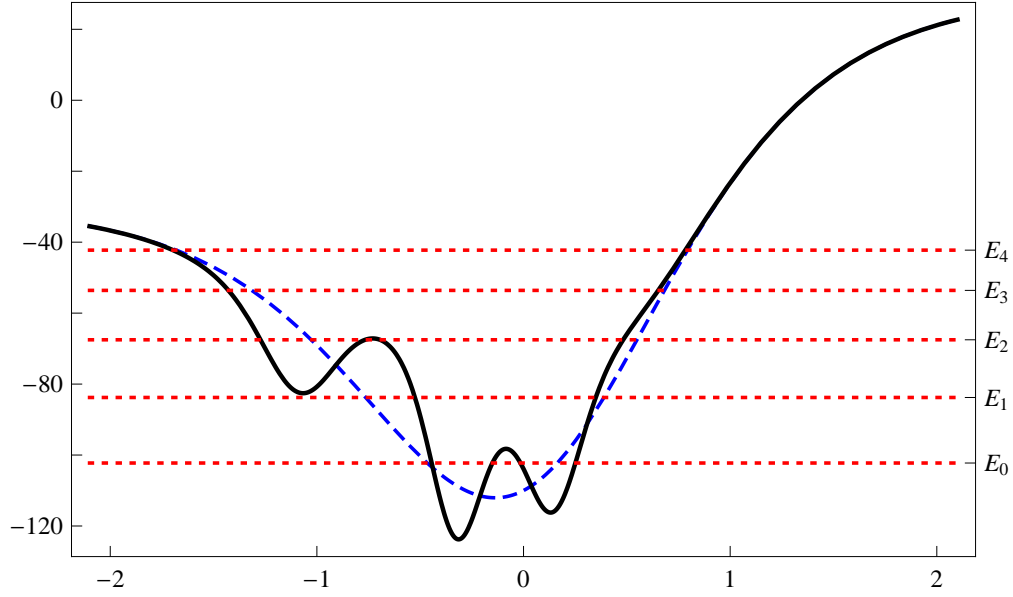}
	\caption{Confluent 2-SUSY partner potential (black solid curve) isospectral to the Rosen Morse potential (blue dashed curve) for $n = 2$, $s = 10$, $\Lambda = 15$ and $w_0 = - \frac{1}{10}$. The five lowest energy levels $E_0=-102.25, \ E_1=-83.78, \ E_2=-67.52, \ E_3=-53.59, \ E_4=-42.25$ of the common spectrum are also shown as the red dotted horizontal lines.}
	\label{fig1}
\end{figure}

\begin{figure}
	\centering
	\includegraphics[width=0.8\textwidth]{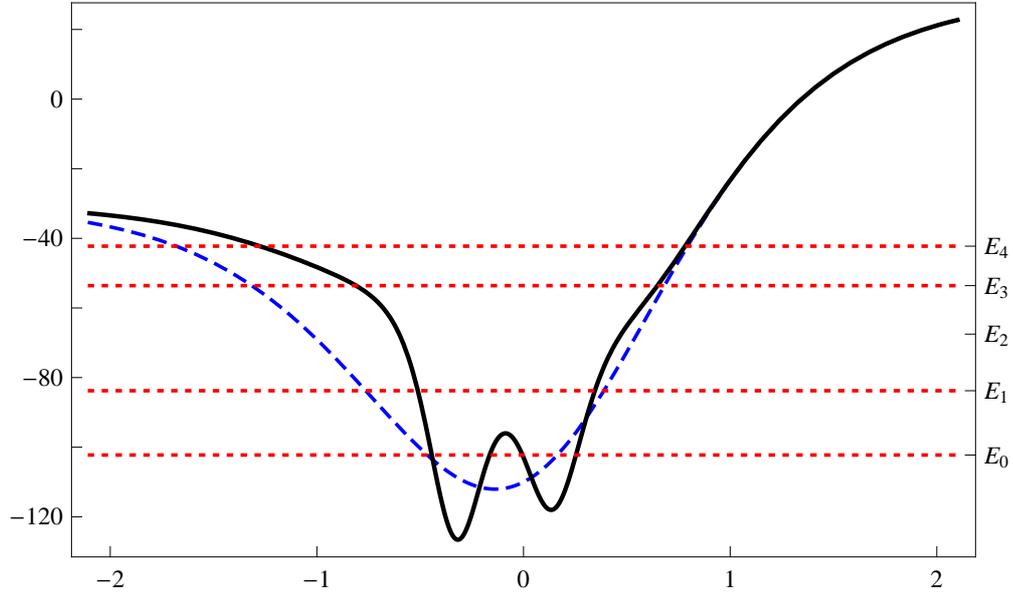}
	\caption{Confluent 2-SUSY partner potential (black solid curve) generated from the Rosen Morse potential with $s = 10$, $\Lambda = 15$ 
	(blue dashed curve) when deleting the eigenvalue  $E_2=-67.52$ (by taking $w_0 = 0$). The four lowest energy levels $E_0=-102.25, \ E_1=-83.78, \ E_3=-53.59, \ E_4=-42.25$ of the new Hamiltonian are also shown as the red dotted horizontal lines.}
	\label{fig2}
\end{figure}

\noindent (b) Deleting a level\\
Departing from the previous results it is possible to delete the level $E_n$, if the parameter $w_0$ takes any of the two
edge values $w_0=0$ or $w_0=1$ defining the border between the non-singular and singular transformations. In both cases the 
transformed bound states associated to $E_m$ for $m\neq n$ become bound states of $\tilde H$, but the wavefunction $\tilde\psi_n \propto \psi_n(x)/w(x)$ associated to $E_n$ is non-normalizable, which implies that $E_n$ does not belong to the spectrum of $\tilde H$.
In particular, for $w_0=0$ and $n=2$ the new potential is shown in Figure \ref{fig2}. It is seen now the disappearance of the 
left local minimum in the new potential as compared with the isospectral case of Fig.~\ref{fig1}, which is due to $w(x)$ now tends to 
zero for $x\rightarrow -\infty$. Something similar will happen to the right when $w_0=1$, since $w(x)$ will tend to zero for 
$x \rightarrow \infty$. It is evident from Fig.~2 that the energy level $E_2=-67.52$ has been deleted. As in the previous isospectral case, the analytic and numerical calculations produce exactly the 
same plot for the new potential.

\noindent (c) Creating a new level\\
Let us choose now $\mathbb{R} \ni \epsilon \neq E_i$, for which the two seed solutions of Eqs.~(\ref{urightnullrm}) and (\ref{uleftnullrm}) can be used. The calculation of the integral of equation (\ref{wofx}) with $x_0 = \infty$ leads to:
\begin{eqnarray}
& \hskip-3.8cm  \int_{x}^{\infty} u_{1}^{2}(y)dy = 2^{\beta -1} \sum\limits_{n=0}^{\infty}\frac{(-s+\frac{\alpha + \beta}{2})_n (s+1+\frac{\alpha+\beta}{2})_n}{(1+\alpha)_{n} n! 2^n} 
\sum\limits_{n'=0}^{\infty}\frac{(-s+\frac{\alpha + \beta}{2})_{n'}(s+1+\frac{\alpha+\beta}{2})_{n'}}{(1+\alpha)_{n'} n'!2^{n'}}  \nonumber \\
& \hskip2.0cm \frac{1}{\alpha+n+n'}(1-\tanh x)^{n+n'+\alpha} \, {}_2F_1\left(n+n'+\alpha,1-\beta,n+n'+\alpha+1,\frac{1-\tanh x}{2}\right). \label{integralrosmor}
\end{eqnarray}

\begin{figure}
	\centering
	\includegraphics[width=0.8\textwidth]{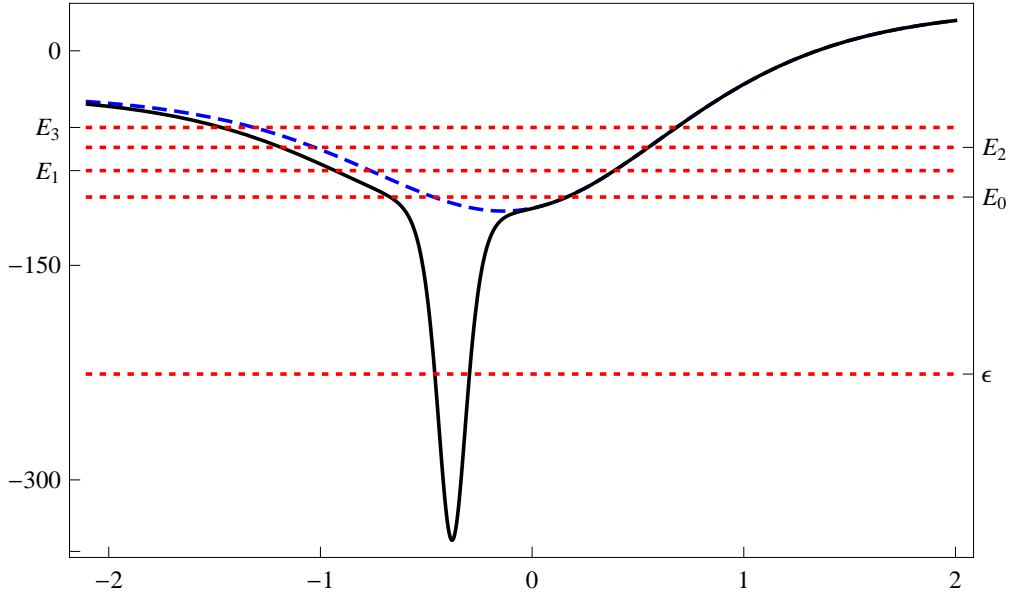}
	\caption{The confluent 2-SUSY partner potential (black solid curve) which arises from the Rosen Morse potential (blue dashed curve) 
	by creating a new level at $\epsilon = -226$ with $s = 10$, $\Lambda = 15$, $w_0 = 0.1$. The five lowest energy levels 
	$\epsilon=-226, \ E_0=-102.25, \ E_1=-83.78, \ E_2=-67.52, \ E_3=-53.59$ of the new Hamiltonian are also shown as the red dotted horizontal lines.}
	\label{fig3}
\end{figure}
With the above expression it is straightforward to calculate the new potential and compare it with the initial one. An example
can be seen in Figure~\ref{fig3}, where equation (\ref{integralrosmor}) was used for $s = 10$, $\Lambda = 15$, 
$w_0 = 0.1$, $\epsilon=-226$ and thus $\alpha=16, \ \beta=-14$. It is worth to stress that for such parameters the hypergeometric 
function in $u_1(x)$ becomes a $9th$ degree polynomial in its argument, thus the two infinite sums of Eq.~(\ref{integralrosmor}) 
truncate at $n=n'=9$. Figure~3 shows also that the energy level $\epsilon=-226$ has been created in order to generate the new potential. 
Once again, the numerically calculated potential turned out to be exactly the same as the generated through our analytic formulas. 

2. {\bf Eckart potential.} \\
Let us consider now the Eckart potential:

\begin{equation}\label{eckpot}
\hskip-1.5cm V(x) = A(A-1){\rm cosech}^2(x) - 2B\coth(x), \quad A>1, \quad B>A^2, \quad 0\leq x < \infty.
\end{equation}
Two linearly independent solutions $\psi^{\pm}$ of the Schr\"odinger equation
\begin{equation}
\frac{d^2 \psi}{dx^2} + [\epsilon - V(x)]\psi = 0,
\end{equation}
where $\epsilon$ is the factorization energy and $V(x)$ is given by Eq.~(\ref{eckpot}), are
\begin{eqnarray}
&& \hskip-0.5cm {\psi}^{+} \!\approx\! (\coth x\!-\!1)^{\frac{\alpha}{2}}~ (1\!+\!\coth x)^{\frac{\beta}{2}} \, {}_2F_{1}\left(A \!+\! \frac{\alpha \!+ \!\beta}{2}, 
1\!-\!A\!+\!\frac{\alpha\!+\!\beta}{2}, 1\!+\!\alpha, \frac{1\!-\!\coth x}{2}\right), \label{psipluseck} \\
&& \hskip-0.5cm {\psi}^{-} \!\approx\! (\coth x\!-\!1)^{\frac{-\alpha}{2}}~ (1\!+\!\coth x)^{\frac{\beta}{2}} \, {}_2F_{1}\left(A \!+ \!\frac{\beta \!- \!\alpha}{2}, 1\!-\!A\!+\!\frac{\beta\!-\!\alpha}{2}, 1\!-\!\alpha, \frac{1\!-\!\coth x}{2}\right), \label{psiminuseck}
\end{eqnarray}
where $\alpha$ and $\beta$ are two new parameters, specified in terms of $\epsilon$ and $B$ as
\begin{equation}
B + \left(\frac{\alpha - \beta}{2}\right)\left(\frac{\alpha + \beta}{2}\right) = 0, \qquad \epsilon + \left(\frac{\alpha + \beta}{2}\right)^2 + \left(\frac{\alpha - \beta}{2}\right)^2 = 0 .
\end{equation}
The normalizable solutions are expressed in terms of Jacobi polynomials as follows:
\begin{equation}\label{boundstateseckart}
\psi_{n}(x) = c_{n} (\coth x-1)^{\frac{\alpha_{n}}{2}} (1 + \coth x)^{\frac{\beta_{n}}{2}} P_n^{\alpha_{n},\beta_{n}}(\coth x),
\end{equation}
where $c_{n}$ is a normalization constant, $\alpha_{n} = -(A + n) + \frac{B}{A + n}, \ \beta_{n} = -(A + n) - \frac{B}{A + n}$, and
\begin{equation}
E_{n} = -(A + n)^2 - \frac{B^2}{(A+n)^2} .
\end{equation}

\begin{figure}
	\centering
	\includegraphics[width=0.8\textwidth]{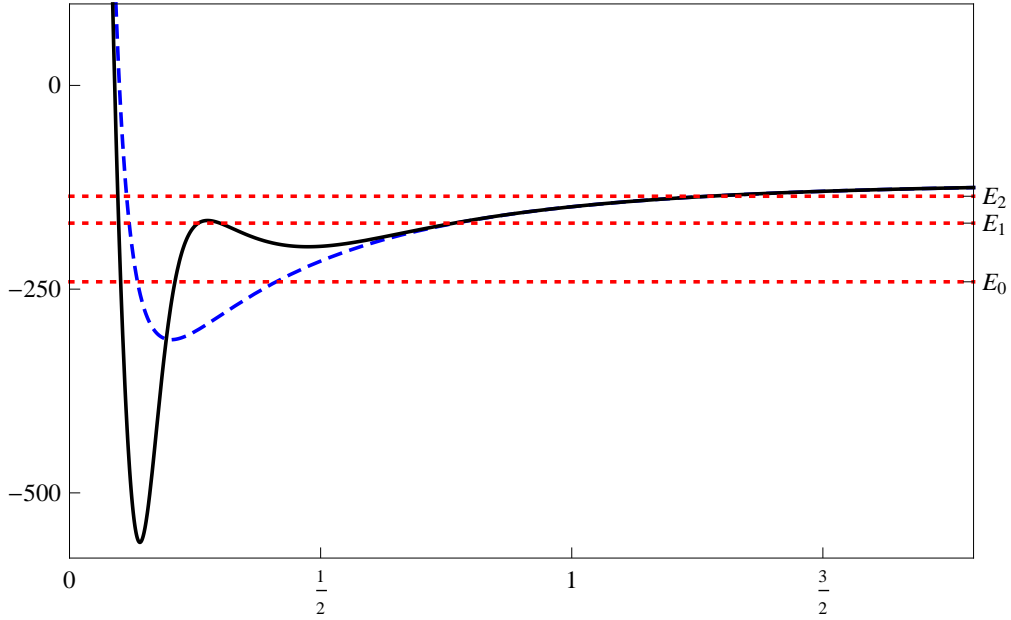}
	\caption{The confluent second-order SUSY partner potential (black solid curve) isospectral to the Eckart potential with $A = 4$, $B = 60$ 
	(blue dashed curve) generated by using the eigenfunction (\ref{boundstateseckart}) for $ n = 0$ and $w_0 = - 0.1$. 
	The three lowest energy levels $E_0=-241, \ E_1=-169, \ E_2=-136$ of the common spectrum are shown as the red dotted horizontal lines.}
	\label{fig4}
\end{figure}

{\bf Spectral design}\\
(a) Isospectral transformation\\
For the confluent isospectral transformation, the normalized solution $\psi_n(x)$ of Eq.~(\ref{boundstateseckart}) with 
$\epsilon = E_n$ is used as seed.  The integral in Eq.~(\ref{wofx}) for $x_0=0$ turns out to be  
\begin{equation}
\int_{0}^{x} \psi_n^2(y)dy = \frac{S_n(x)}{S_n(\infty)},
\end{equation}
where
\begin{equation}
\begin{array}{l}
\hskip-0.1cm S_n(x) \!=\!\! \sum\limits_{m'=0}^n \! \left(\!\!\!\begin{array}{c} n\!+\!\alpha_n\\m'\end{array}\!\!\!\right)\!\left(\!\!\!\begin{array}{c} n\!+\!\beta_n\\n-m'\end{array}\!\!\!\right) \frac{1}{(-1)^{m'} 2^{m'}} \sum\limits_{m=0}^n \! \left(\!\!\!\begin{array}{c} n\!+\!\alpha_n\\m\end{array}\!\!\!\right)\!\left(\!\!\!\begin{array}{c} n\!+\!\beta_n\\n-m\end{array}\!\!\!\right) \frac{1}{(-1)^{m} 2^{m}} \frac{1}{(\beta_n+m+m')}\\
\hskip1.8cm (1\!+\!\coth x)^{\beta_n+m+m'} \, {}_2F_1(\beta_n\!+\!m\!+\!m',-\alpha_n\!-\!2n\!+\!m\!+\!m'\!+\!1,\beta_n\!+\!m\!+\!m'\!+\!1,\frac{1\!+\!\coth x}{2}),
\end{array} 
\end{equation}
\begin{equation}
\begin{array}{l}
\hskip-1.0cm S_n(\infty) \!= 2^{\beta_n} \sum\limits_{m'=0}^n \! \left(\!\!\!\begin{array}{c} n+\alpha_n\\m'\end{array}\!\!\!\right)\left(\!\!\!\begin{array}{c} n+\beta_n\\n-m'\end{array}\!\!\!\right) \frac{1}{(-1)^{m'}} \sum\limits_{m=0}^n \! \left(\!\!\!\begin{array}{c} n+\alpha_n\\m\end{array}\!\!\!\right)\left(\!\!\!\begin{array}{c} n+\beta_n\\n-m\end{array}\!\!\!\right) \frac{1}{(-1)^{m}}\\
\hskip2.2cm \frac{1}{\beta_n+m+m'} \frac{\Gamma (\beta_n + m + m'+1) \Gamma(\alpha_n+2n-m-m')}{\Gamma(\alpha_n+\beta_n+2n)} .
\end{array}
\end{equation}
In this case $w(x)$ will be nodeless if $w_0 \in (-\infty,0) \cup  (1,\infty)$.
The confluent supersymmetric partner potential is given by
\begin{equation}
\tilde{V}(x) = A(A-1){\rm cosech}^2(x) - 2B\coth(x)  - 2\left[\frac{w^{\prime}(x)}{w(x)}\right]^{\prime} .
\end{equation}
Since $\tilde{\psi}_{n}(x) \propto \frac{\psi_n(x)}{w(x)}$ satisfies
\begin{equation}
\lim_{x\rightarrow 0}\tilde{\psi}_{n}(x) = \lim_{x\rightarrow \infty}\tilde{\psi}_{n}(x) = 0 ,
\end{equation}
then $\epsilon = E_n \in {\rm Sp}(\tilde{H})$, and thus $H$ and $\tilde{H}$ are isospectral.\\
An example of the new and initial potentials (black continuous and blue dashed lines, respectively) when the seed is the ground state
eigenfunction $\psi_0(x)$ is shown in Fig.~\ref{fig4} for $w_0=-0.1, \ A=4, \ B=60$. 
The first three energy levels are also shown in the same Figure. The numerical integration of $w(x)$ produced exactly the same plot for the new potential.

\begin{figure}
	\centering
	\includegraphics[width=0.8\textwidth]{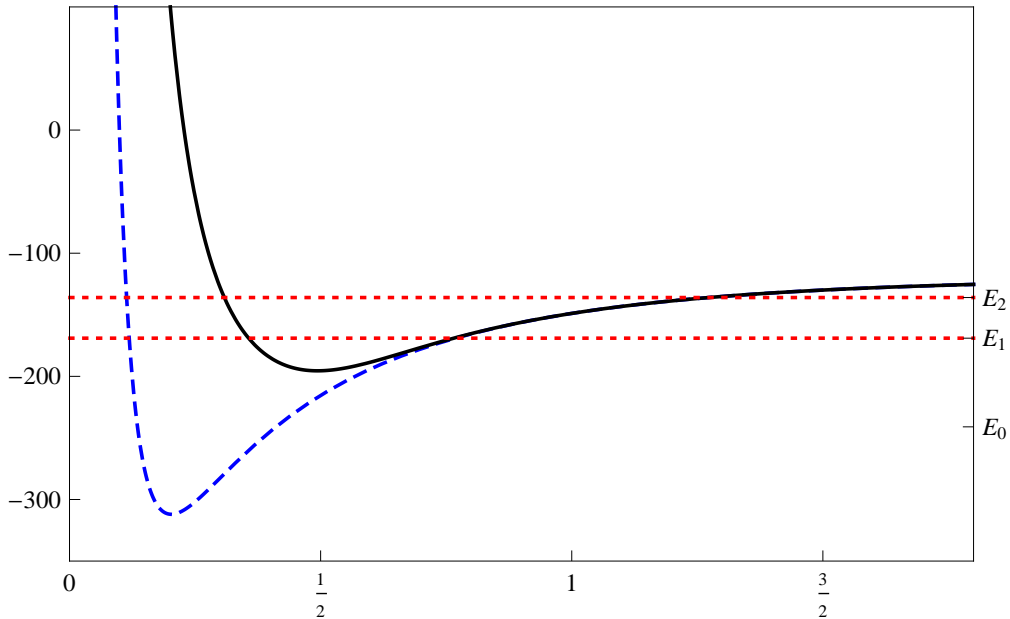}
	\caption{The confluent second-order SUSY partner potential (black solid curve) generated from the Eckart potential with $A = 4$, $B = 60$ 
	(blue dashed curve) by deleting the energy level $E_0=-241$ when taking $w_0 = 0$. The two lowest energy levels $E_1=-169, \ E_2=-136$ of the new Hamiltonian are also drawn as the red dotted horizontal lines.}
	\label{fig5}
\end{figure}

\noindent (b) Deleting the level $E_n$.\\
If the parameter $w_0$ of the previous confluent transformation takes any of the two border values $w_0=0$ or $w_0=1$ this 
transformation is still non-singular, mapping the bound states of $H$ into bound states of $\tilde H$ for $m\neq n$, but the
formal eigenfunction $\tilde\psi_n \propto \psi_n(x)/w(x)$ of $\tilde H$ associated to $E_n$ leaves to be square-integrable.
This means that the energy levels of $\tilde H$ coincide with those of $H$ except by $E_n$, which has been deleted. In order 
to illustrate this case we have taken $w_0=0$ in our previous formulas, which implies that
$$
w(x) = -\int_0^x \psi_n^2(y) dy = - \frac{S_n(x)}{S_n(\infty)}.
$$
The corresponding initial and final potentials for $n=0$ are shown in Figure~\ref{fig5}. As it is seen, the final potential does not tend to
the initial one for $x\rightarrow 0$, since in this limit $w(x) \rightarrow 0$ which induces a change in the coefficient of the 
singularity of the initial Eckart potential at $x=0$. Comparison of Fig.~4 and Fig.~5 clearly shows the deletion of the level $E_0=-241$. 
Once again, the numerical calculation led to the same plot for the new potential as the analytical result.

\noindent(c) Creating a new level\\
Let us choose now $\mathbb{R}\ni \epsilon \neq E_n$. In order that the argument $\frac{1-\coth x}{2}$ of the Hypergeometric function in 
Eqs.~(\ref{psipluseck},\ref{psiminuseck}) will fall completely within the domain of convergence of the Gauss Hypergeometric series ${}_2F_1(a,b,c,z) = \sum\limits_{n=0}^{\infty} \frac{(a)_n(b)_n}{(c)_n} \frac{z^n}{n!}$ the following transformation formula is used \cite{abra}:
\begin{equation}
{}_2F_1(a,b,c,z) = (1-z)^{-b} \, {}_2F_1\left(b,c-a,c, \frac{z}{z-1}\right)
\end{equation}
in the two linearly independent solutions of Eqs.~(\ref{psipluseck},\ref{psiminuseck}), leading to:
\begin{eqnarray}
&& \hskip-0.5cm \psi^+ \approx e^{-\alpha x} (1-e^{-2x})^{1-A} \, {}_2F_{1}\left(1-A+\frac{\alpha+\beta}{2}, 1-A+\frac{\alpha-\beta}{2}, 1+\alpha, e^{-2x}\right), \\
&& \hskip-0.5cm \psi^- \approx e^{\alpha x} (1-e^{-2x})^{1-A} \, {}_2F_{1}\left(1-A+\frac{\beta-\alpha}{2}, 1-A-\frac{\alpha+\beta}{2}, 1-\alpha, e^{-2x}\right).
\end{eqnarray} 
The seed solutions satisfying the right asymptotic behavior, vanishing for ${x\rightarrow\infty}$ or ${x\rightarrow 0}$ respectively, are
\begin{eqnarray}
&& \hskip-1.8cm u_{1} (x) =  e^{-\alpha x} (1-e^{-2x})^{1-A} \, {}_2F_{1}\left(1-A+\frac{\alpha+\beta}{2}, 1-A+\frac{\alpha-\beta}{2}, 1+\alpha, e^{-2x}\right), \label{uplus}\\
&& \hskip-1.8cm u_{2} (x) =  e^{-\alpha x} (1-e^{-2x})^{1-A} \, {}_2F_{1}\left(1-A+\frac{\alpha+\beta}{2}, 1-A+\frac{\alpha-\beta}{2}, 1+\alpha, e^{-2x}\right) \nonumber \\
&& \hskip-0.4cm - D^- e^{\alpha x} (1-e^{-2x})^{1-A} \, {}_2F_{1}\left(1-A+\frac{\beta-\alpha}{2}, 1-A-\frac{\alpha+\beta}{2}, 1-\alpha, e^{-2x}\right),
\end{eqnarray}
where $D^- = \frac{\Gamma (1+\alpha)\Gamma (A - \frac{\alpha-\beta}{2})\Gamma (A-\frac{\alpha+\beta}{2})}{\Gamma (1-\alpha)\Gamma (A + \frac{\alpha+\beta}{2})\Gamma (A+\frac{\alpha-\beta}{2})}, \ \alpha = \sqrt{-(\epsilon + 2B)}, \ \beta^2 = -\epsilon + 2B$,
$\epsilon$ being the factorization energy. Note that $\beta$ and $\alpha$ will be real for $2B > \epsilon$ and $\epsilon+ 2B < 0$, respectively.\\

\begin{figure}
	\centering
	\includegraphics[width=0.8\textwidth]{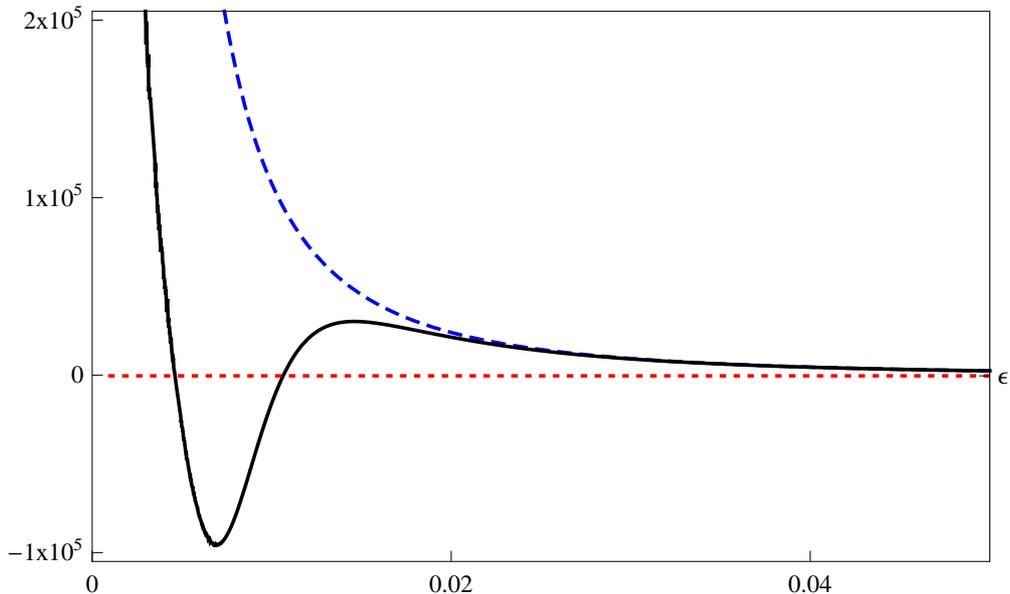}
	\caption{The confluent 2-SUSY partner potential (black solid curve) which arises from creating a new level at $\epsilon = -409$ 
	for the Eckart potential with $A = 4$, $B = 60$ (blue dashed curve) and $w_0 = 0.01$. The red dotted horizontal line represents 
	the new energy level.}
	\label{fig6}
\end{figure}

Let us take now the seed solution $u_1(x)$ of Eq.~(\ref{uplus}), which vanishes when $x\rightarrow \infty$. The
evaluation of the integral in Eq.~(\ref{wofx}) with $x_0 = \infty$ leads to
\begin{eqnarray}
&& \hskip-1.0cm \int_{\infty}^{x} \! u_{1}^{2}(y)dy \!=\! -\frac{1}{2} \! \sum\limits_{n=0}^{\infty} \!\! \frac{(1-A+\frac{\alpha + \beta}{2})_{n} (1-A+\frac{\alpha-\beta}{2})_{n}}{(1+\alpha)_{n} n!} \!
\sum\limits_{n'=0}^{\infty} \!\! \frac{(1-A+\frac{\alpha + \beta}{2})_{n'} (1-A+\frac{\alpha-\beta}{2})_{n'}}{(1+\alpha)_{n'} n'!}  \nonumber \\
&& \hskip9.0cm B(e^{-2x}, \alpha \!+\! n \!+ \!n', 3\!-\!2A), \label{usqeck}
\end{eqnarray}
where $B(z,a,b)$ is the incomplete Beta function \cite{abra}.
The confluent second-order SUSY partner potentials $\tilde V(x)$ and $V(x)$ are illustrated in Figure~\ref{fig6} for 
$A=4, \ B=60$, $w_0 = 0.01$, $\epsilon=-409$, and thus $\alpha=17, \ \beta=-23$. Once again,  
for such parameters the hypergeometric function appearing in $u_1(x)$ becomes a $6th$ degree polynomial in its argument, 
and thus the two infinite sums of Eq.~(\ref{usqeck}) truncate at $n=n'=6$. It is seen from Fig.~6 that a new energy level at 
$\epsilon=-409$ has been created. As in the previous examples, the analytic and numeric calculations produce the same plots
for the new potential. 

\section{Conclusion}
In this article the confluent second order supersymmetric partners of the Rosen Morse II and Eckart potentials are studied. Several possible spectral modifications are produced by taking appropriate seed solutions and associated factorization energies. For real factorization energies, it is possible to generate the partner potentials: (a) with identical spectrum as of the original potential; (b) with one level deleted from the initial spectrum: (c) with one new level embedded into the spectrum of the original potential.\\ As for physical applications of these types of SUSY transformed potentials, it should be mentioned that SUSY transformations have been applied recently for refractive index engineering of optical materials \cite{M10}. In particular, SUSY transformations of optical systems can be used to reduce the refractive index needed in a given structure. This can be done through a hierarchical ladder of superpartners obtained by sequentially removing the bound states. On the other hand, SUSY transformations which add modes to a given structure are used to locally increase the permittivity of an optical material. For a detailed discussion, we refer the reader to \cite{M10}.\\ One of the interesting issues to be addressed in future work is to consider spectral modification of rationally extended potentials, whose bound state wavefuctions are associated with exceptional orthogonal polynomials \cite{quesne10, quesne20}.


\end{document}